\documentclass{elsart}
\usepackage{epsfig}
\usepackage{color}

\newcommand{\affuni}[2]{Dipartimento di Fisica dell'Universit\`a #1, #2, Italy.}

\newcommand{\affinfnm}[2]{INFN Sezione di #2, #2, Italy.}
\newcommand{\affinfnn}[2]{INFN Sezione #1, #2, Italy.}

\begin{document}

\begin{frontmatter}

\title{Search for the decay $\phi\rightarrow K^0\overline{K}^0\gamma$ with the KLOE experiment}

\collab{The KLOE Collaboration}
\author[Na,infnNa]{F.~Ambrosino},
\author[Frascati]{A.~Antonelli},
\author[Frascati]{M.~Antonelli},
\author[Roma2,infnRoma2]{F.~Archilli},
\author[Mainz]{P.~Beltrame},
\author[Frascati]{G.~Bencivenni},
\author[Frascati]{S.~Bertolucci},
\author[Roma1,infnRoma1]{C.~Bini},
\author[Frascati]{C.~Bloise},
\author[Roma3,infnRoma3]{S.~Bocchetta},
\author[Frascati]{F.~Bossi},
\author[infnRoma3]{P.~Branchini},
\author[Frascati]{G.~Capon},
\author[Frascati]{T.~Capussela},
\author[Roma3,infnRoma3]{F.~Ceradini},
\author[Frascati]{P.~Ciambrone},
\author[Frascati]{E.~De~Lucia},
\author[Roma1,infnRoma1]{A.~De~Santis},
\author[Frascati]{P.~De~Simone},
\author[Roma1,infnRoma1]{G.~De~Zorzi},
\author[Mainz]{A.~Denig},
\author[Roma1,infnRoma1]{A.~Di~Domenico},
\author[infnNa]{C.~Di~Donato},
\author[Roma3,infnRoma3]{B.~Di~Micco},
\author[Frascati]{M.~Dreucci},
\author[Frascati]{G.~Felici},
\author[Roma1,infnRoma1]{S.~Fiore\corauthref{cor}},
\ead{salvatore.fiore@roma1.infn.it}
\corauth[cor]{Corresponding author. Address: p.le Aldo Moro, 2 - 00185 Roma - Italy. Phone/Fax +390649913429}
\author[Roma1,infnRoma1]{P.~Franzini},
\author[Frascati]{C.~Gatti},
\author[Roma1,infnRoma1]{P.~Gauzzi},
\author[Frascati]{S.~Giovannella},
\author[infnRoma3]{E.~Graziani},
\author[Frascati]{G.~Lanfranchi},
\author[Frascati,StonyBrook]{J.~Lee-Franzini},
\author[Frascati,Energ]{M.~Martini},
\author[Na,infnNa]{P.~Massarotti},
\author[Na,infnNa]{S.~Meola},
\author[Frascati]{S.~Miscetti},
\author[Frascati]{M.~Moulson},
\author[Mainz]{S.~M\"uller},
\author[Frascati]{F.~Murtas},
\author[Na,infnNa]{M.~Napolitano},
\author[Roma3,infnRoma3]{F.~Nguyen},
\author[Frascati]{M.~Palutan},
\author[infnRoma1]{E.~Pasqualucci},
\author[infnRoma3]{A.~Passeri},
\author[Frascati,Energ]{V.~Patera},
\author[Frascati]{P.~Santangelo},
\author[Frascati]{B.~Sciascia},
\author[Frascati]{T.~Spadaro},
\author[Roma1,infnRoma1]{M.~Testa},
\author[infnRoma3]{L.~Tortora},
\author[infnRoma1]{P.~Valente},
\author[Frascati]{G.~Venanzoni},
\author[Frascati,Energ]{R.Versaci},
\author[Frascati,Beijing]{G.~Xu}\\


\address[Frascati]{Laboratori Nazionali di Frascati dell'INFN, 
Frascati, Italy.}
\address[Mainz]{Institut f\"ur Kernphysik, 
Johannes Gutenberg - Universit\"at Mainz, Germany.}
\address[Na]{Dipartimento di Scienze Fisiche dell'Universit\`a 
``Federico II'', Napoli, Italy}
\address[infnNa]{INFN Sezione di Napoli, Napoli, Italy}
\address[Energ]{Dipartimento di Energetica dell'Universit\`a 
Sapienza di Roma, Roma, Italy.}
\address[Roma1]{\affuni{Sapienza di Roma}{Roma}}
\address[infnRoma1]{\affinfnm{``La Sapienza''}{Roma}}
\address[Roma2]{\affuni{``Tor Vergata''}{Roma}}
\address[infnRoma2]{\affinfnn{Roma Tor Vergata}{Roma}}
\address[Roma3]{\affuni{``Roma Tre''}{Roma}}
\address[infnRoma3]{\affinfnn{Roma Tre}{Roma}}
\address[StonyBrook]{Physics Department, State University of New 
York at Stony Brook, USA.}
\address[Beijing]{Institute of High Energy Physics of Academica Sinica, Beijing, China.}


\begin{abstract}
We have searched for the decay $\phi\rightarrow K^0\overline{K}^0\gamma$ , by detecting $K_S$ pairs plus a photon and with the $K_S$-mesons decaying to $\pi^+\pi^-$, in a sample of about 1.5$\times 10^9$ $\phi$-decays collected by the KLOE experiment at DA$\Phi$NE. The reaction proceeds through the intermediate states ${f_0(980)}\gamma $, ${a_0(980)}\gamma $.
We find five events with 3.2 events expected from
background processes. We obtain the upper limit: 
$BR(\phi \to K^0\overline{K}^0\gamma) < 1.9 \times 10^{-8}$ at 90\% C.L. .
\end{abstract}

\end{frontmatter}

\section{Introduction}
We have searched for the decay 
$\phi\rightarrow K^0\overline{K}^0\gamma$ with the KLOE experiment at DA$\Phi$NE. This
decay has never been observed. In the decay the $K^0\overline{K}^0$
pair is produced with quantum numbers $J^{PC}=0^{++}$ 
, so that the two-kaon state is symmetric, and can be written in terms of $K_S, K_L$ as
\begin{equation}
|K^0\overline{K}^0>={{|K_SK_S>-|K_LK_L>}\over{\sqrt{2}}} .
\label{stato}
\end{equation}

The $K^0\overline{K}^0$ state can be in both singlet and triplet isospin state, so that the decay proceeds mainly via intermediate $f_0$(980) ($I$=0) and $a_0$(980) ($I$=1) scalar mesons, 
\begin{equation}
\phi\rightarrow ({\rm f_0(980)+a_0(980)})\gamma\rightarrow K^0\overline{K}^0\gamma . 
\label{scheme}
\end{equation}

Thus, a measurement of the $\phi\rightarrow K^0\overline{K}^0\gamma$ branching ratio (BR) can give information on the structure of light scalar mesons. The evaluation of the branching ratio depends on the
way the scalar field dynamics is introduced and on the size of the couplings of the scalar mesons to the kaons. Moreover, interference between the f$_0$ and a$_0$ amplitudes may also contribute to the $\phi\rightarrow K^0\overline{K}^0\gamma$ decay amplitude. Predictions found in literature are mainly based on the decay scheme (\ref{scheme}), and give values 
ranging over a large interval, from few $\times 10^{-9}$ to $10^{-7}$ ~\cite{Achasov1,Fajfer,Lucio,Bramon2,Oller1,Achasov2,Nussinov,Oller2,Escribano,Gokalp}. In this respect, a similar decay into charged kaons, $\phi \to K^+K^-\gamma$, gives less information since the scalar dynamics is masked by the final state radiation.


The signature of the radiative decay of the $\phi$ meson into the state (\ref{stato}) is the presence of either two
$K_S$ or two $K_L$ (and a low energy photon). 
$KK$ invariant mass is limited 
by the $\phi$ mass (1020 MeV) and 
by twice the kaon mass
(995 MeV). 
The photon energy is a function of the invariant di-kaon mass $M_{KK}$ and is
given by
\begin{equation}
E_{\gamma}={{m^2_{\phi}-M_{KK}^2}\over{2m_{\phi}}} .
\label{maxene}
\end{equation}
We search for final states with a $K_SK_S$ pair, with both $K_S$ decaying to $\pi^+\pi^-$. This corresponds to a reduction of the observable 
rate of 
\begin{equation}
{{1}\over{2}}\times [BR(K_S\rightarrow \pi^+\pi^-)]^2 \sim 2/9 ,
\end{equation}
where the factor $1/2$ accounts for the choice of the $K_SK_S$ final state 
among the two possibilities for the $K^0\overline{K}^0$ state. 
We search for two decay 
vertices very close to the $e^+e^-$ interaction region, both having two tracks with opposite charge and
invariant mass of the charged secondaries equal to the kaon mass. The invariant mass of the kaon pair must be smaller than the $\phi$ mass. A low energy photon must be detected with a momentum compatible with the kinematics of  the event.  
\par
\begin{figure}
\begin{center}
   \mbox{ \epsfig{figure=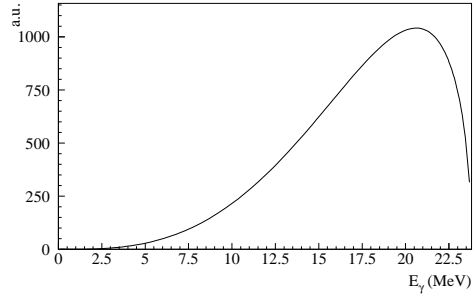,width=0.44\textwidth,clip=,bbllx=50pt,bblly=250pt,bburx=550pt,bbury=570pt}  } 
\end{center} 
\caption{\small Expected $\phi\rightarrow K^0\overline{K}^0\gamma$ photon energy spectrum in the $\phi$ centre-of-mass, according to phase-space and
  radiative decay dynamics. }
\label{spectrum}
\end{figure}

 The expected photon energy spectrum ($E_{\gamma}$) is shown in fig.~\ref{spectrum}, corresponding to a $K_SK_S$ invariant mass ($M_{KK}$) spectrum generated assuming only 
phase-space and radiative decay dynamics: 

\begin{equation}
\frac{d\Gamma}{dM_{KK}}\propto \left({{m^2_{\phi}-M_{KK}^2}\over{2m_{\phi}}}\right)^3 
\sqrt{1- \frac{4m^2_{K0}}{M_{KK}^2}} .
\label{spect_gen}
\end{equation}

The highest rate is expected for events with photon energy close to the maximum value (see eq.~\ref{maxene}). The selection criteria have been optimized in order to maximize the sensitivity to the searched signal~\cite{feldcous}.

\section{The KLOE experiment}
DA$\Phi$NE is an $e^+ e^-$ collider 
running at a centre-of-mass energy $\sqrt{s}$ = m$_\phi$ = 1.02 GeV. The beams collide with a crossing angle of ($\pi$ - 0.025) rad which gives a $\phi$ meson momentum of $\sim$15 MeV (12 MeV for a small part of the dataset) towards the collider centre. The KLOE detector consists of a large-volume cylindrical
drift chamber~\cite{dc} (3.3 m length and 2 m radius), operated
with a 90\% helium - 10\% isobutane gas mixture, surrounded by
a sampling calorimeter~\cite{emc} made of lead and scintillating fibers
providing a solid angle coverage of 98\%. The tracking
chamber and the calorimeter are surrounded by a superconducting
coil that provides an (approximately) axial magnetic field B = 0.52 T, the axis being the bisector of the external angle of the $e^+$ and $e^-$ beams. The
drift chamber has a momentum resolution $\sigma (p_\bot)/p_\bot \sim$ 0.4\%. Photon energies and arrival times are measured in the
calorimeter with resolutions $\sigma_E/E$ = 5.7\%$/\sqrt{E}$(GeV) and
$\sigma_t$ = 57 ps$/\sqrt{E}$(GeV) $\bigoplus$ 100 ps. The trigger~\cite{trigger} is based on
the detection of at least two energy deposits in the calorimeter
above a threshold that ranges between 50 and 150 MeV.

For this search the full data sample acquired at the $\phi$-peak energy by the KLOE experiment between years 2001 and 2005 has been used. After quality selection, a data sample corresponding to an integrated luminosity of 2.18 fb$^{-1}$ is available for the present analysis. The luminosity is measured with an error of less than 1\% by counting large-angle Bhabha scattering events~\cite{lumbha}.

The KLOE Monte Carlo simulation (MC) package, GEANFI~\cite{OFFLINE}, has been used to provide high-statistics
samples. The total number of events in each
sample is generated with an integrated luminosity scale factor (LF) defined as LF~=~integrated luminosity (MC generation)/integrated luminosity (data). This scale factor is different
 for different $\phi$ decay modes and different data taking periods.
  All $\phi$ decays have been simulated with LF=1 or 2, {\it CP}-violating rare decays, such as $K_S K_L \to \pi^+\pi^-\pi^+\pi^-$, have been simulated with LF=10 and 20.
DA$\Phi$NE operating conditions during data taking,
 machine parameters and background are adjusted in the MC on a
run-by-run basis.

 
Simulations of all $\phi$ decays and {\it CP}-violating rare decays have been used 
to evaluate sources of background mimicking $\phi\rightarrow K^0\overline{K}^0\gamma$ decays. The initial-state radiation (ISR) process $e^+e^-\rightarrow \gamma K^0\overline{K}^0$  is taken into account in all the simulations.
A further possible background source 
 is the continuum production of charged 
pions through the process $e^+e^- \rightarrow \pi^+\pi^-\pi^+\pi^-(\gamma)$. 
 A pure 
phase-space generation with LF=10 
has been used to simulate this kind of events. 
The
measured cross section of this 
process in the $\phi$ region~\cite{4pcmd2} has been used to
normalize MC to data.

To simulate the signal we generated a MC sample of 
$10^4$ $\phi \to K_SK_S\gamma\to\pi^+\pi^-\pi^+\pi^-\gamma$ events, developing a modified version of the PHOKHARA5 generator~\cite{rodrigo}. Phokhara is a Monte Carlo event generator which simulates the electron-positron annihilation into hadrons plus an energetic photon from initial state radiation (ISR) process, at the next-to-leading order (NLO) accuracy. This simulation relies on general assumptions for the branching ratio dependence on the photon energy and on phase space, according to eq.~\ref{spect_gen}.

\section{Data analysis}

We search for $K_SK_S\gamma$ events by requiring two pairs of tracks with opposite charge, each pair originating from a vertex contained inside a cylindrical volume centered on the interaction point and the beam line, 16 cm long in z and of 3 cm radius.
We reconstruct the invariant masses $m_1$ and $m_2$, for each pair of tracks, using the pion mass hypothesis. The distribution of signal events in the plane $m_1$, $m_2$ is well contained inside a circle of a few MeV radius centered on the $K_S$ mass. We
require the events to satisfy a 4 MeV cut on this radius.

\begin{figure}
  \begin{center}  
   \mbox{ \epsfig{figure=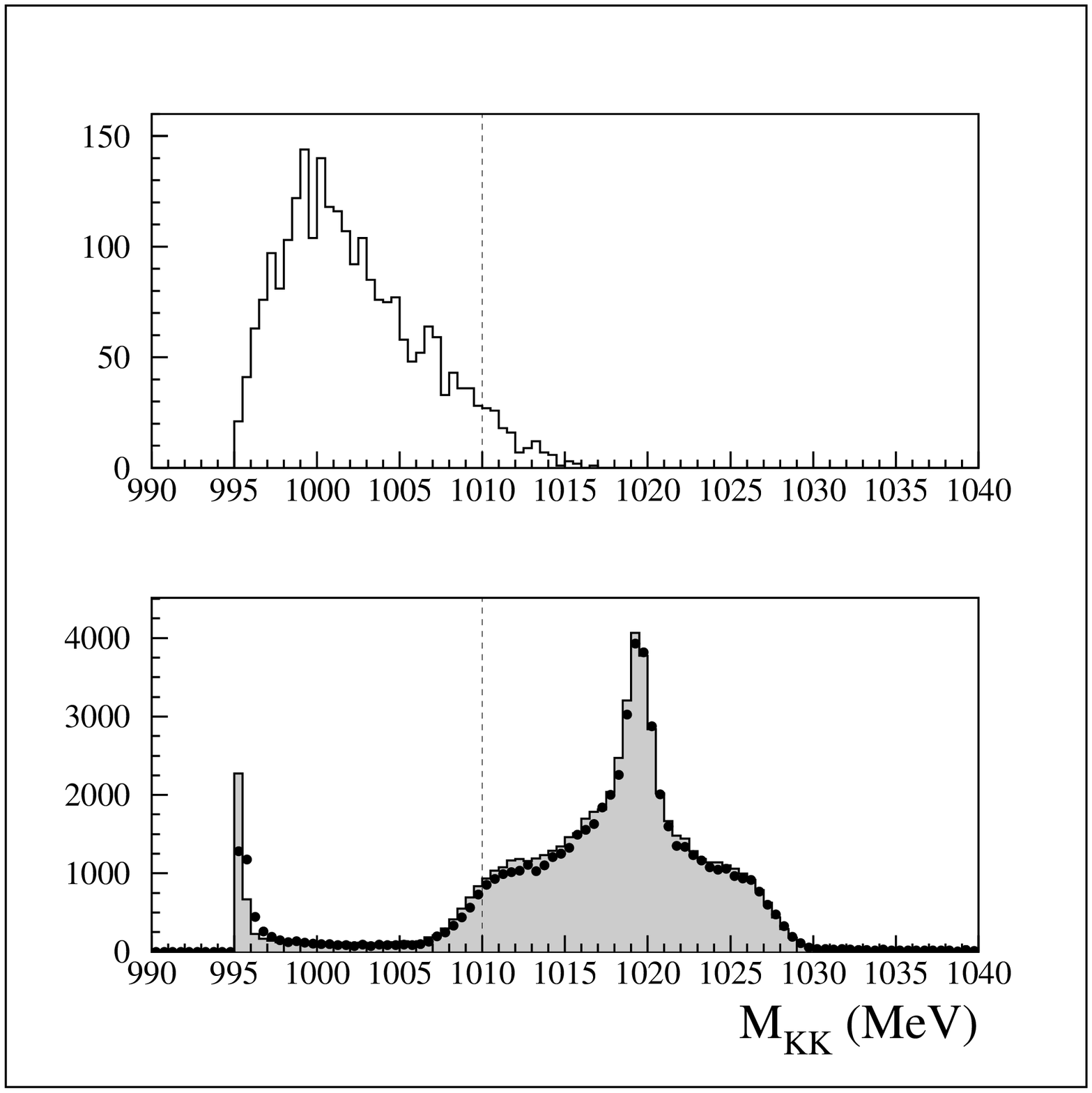,width=0.44\textwidth,clip=,bbllx=50pt,bblly=210pt,bburx=550pt,bbury=700pt}  } 
   \mbox{ \epsfig{figure=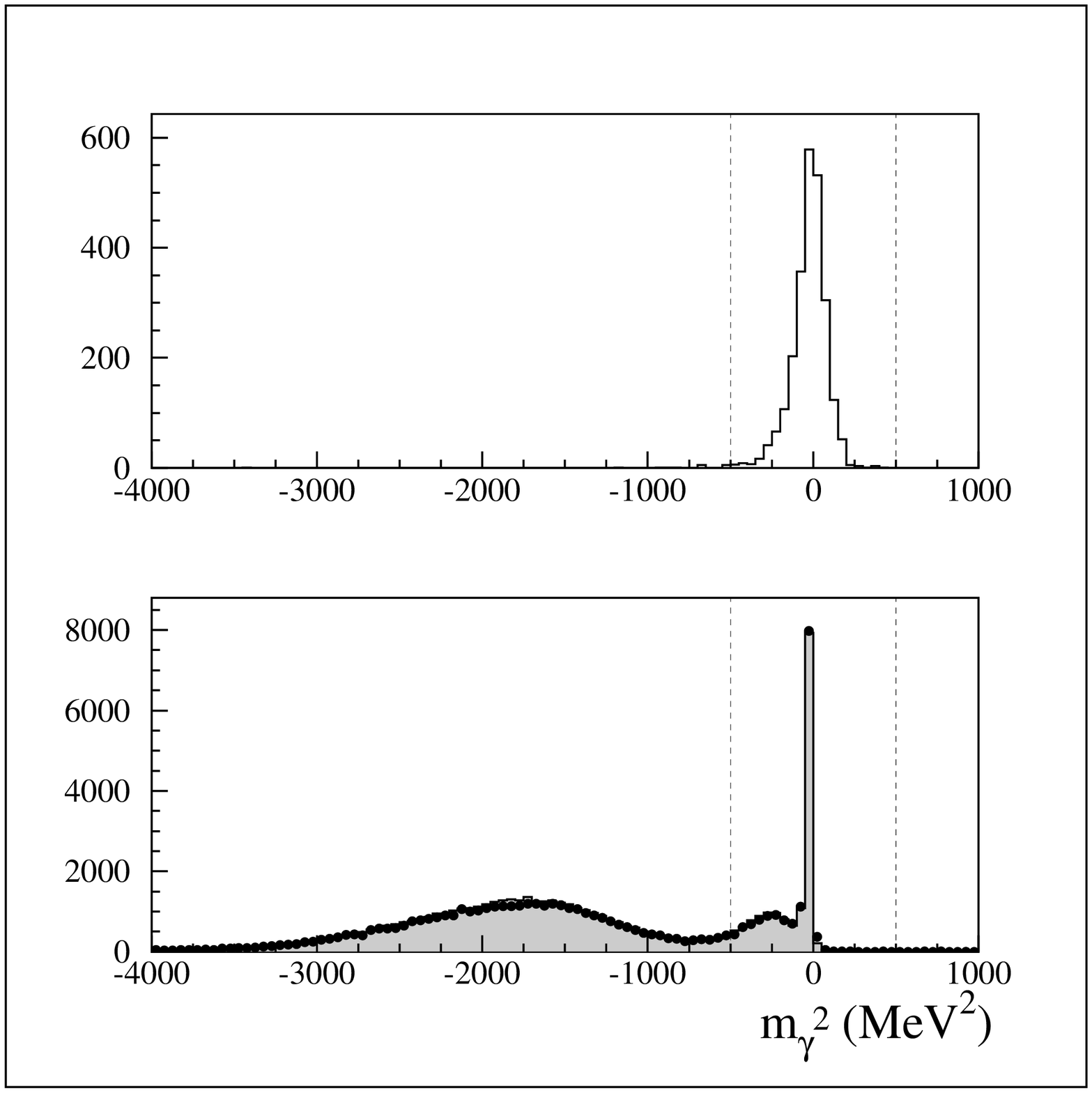,width=0.44\textwidth,clip=,bbllx=50pt,bblly=210pt,bburx=550pt,bbury=700pt}  } 
  \end{center}
\vspace{1.cm}
  \caption{\small Distributions of the two-kaon invariant mass (left) and the $m_{\gamma}^2$ variable (right). Top: $\phi\rightarrow K^0\overline{K}^0\gamma$ MC. Bottom: data (dots) and background MC (grey).
Selection cuts are shown. Background MC samples are normalized to data taking into account the different LF values.
}
   \label{fig:sidebox}
\end{figure} 

Once the two $K_S$ mesons are identified, we build the invariant mass of the $K_S$ pair, $M_{KK}$. In order to reject background from $\phi \to K_SK_L \to \pi^+ \pi^- \pi^+ \pi^-$ events, we retain only those events with $K_SK_S$ invariant mass below 1010 MeV.
Moreover, in order to reject badly reconstructed $K_S$ semileptonic decays, which could mimic a two-pion decay, we define the missing mass $m_\gamma$  using four-momentum conservation:
\begin{equation}
m_\gamma^2 = (P_\phi - P_{K_{S1}} - P_{K_{S2}})^2 ,
\end{equation}
where $P_\phi$, $P_{K_{S1}}$ and $P_{K_{S2}}$ are the 4-vectors of the $\phi$ and of the two reconstructed $K_S$. This variable must be zero in case of two $K_{S,L}$ to $\pi^+ \pi^-$ decays, even with an extra photon, and is required to be zero within $\pm$500 MeV$^2$.
Figure \ref{fig:sidebox} shows the distribution of signal simulation, data and background simulation for  $M_{KK}$ and $M_\gamma^2$ . Ninety five events survive these cuts.


\begin{figure}
  \begin{center}  
   \mbox{\epsfig{figure=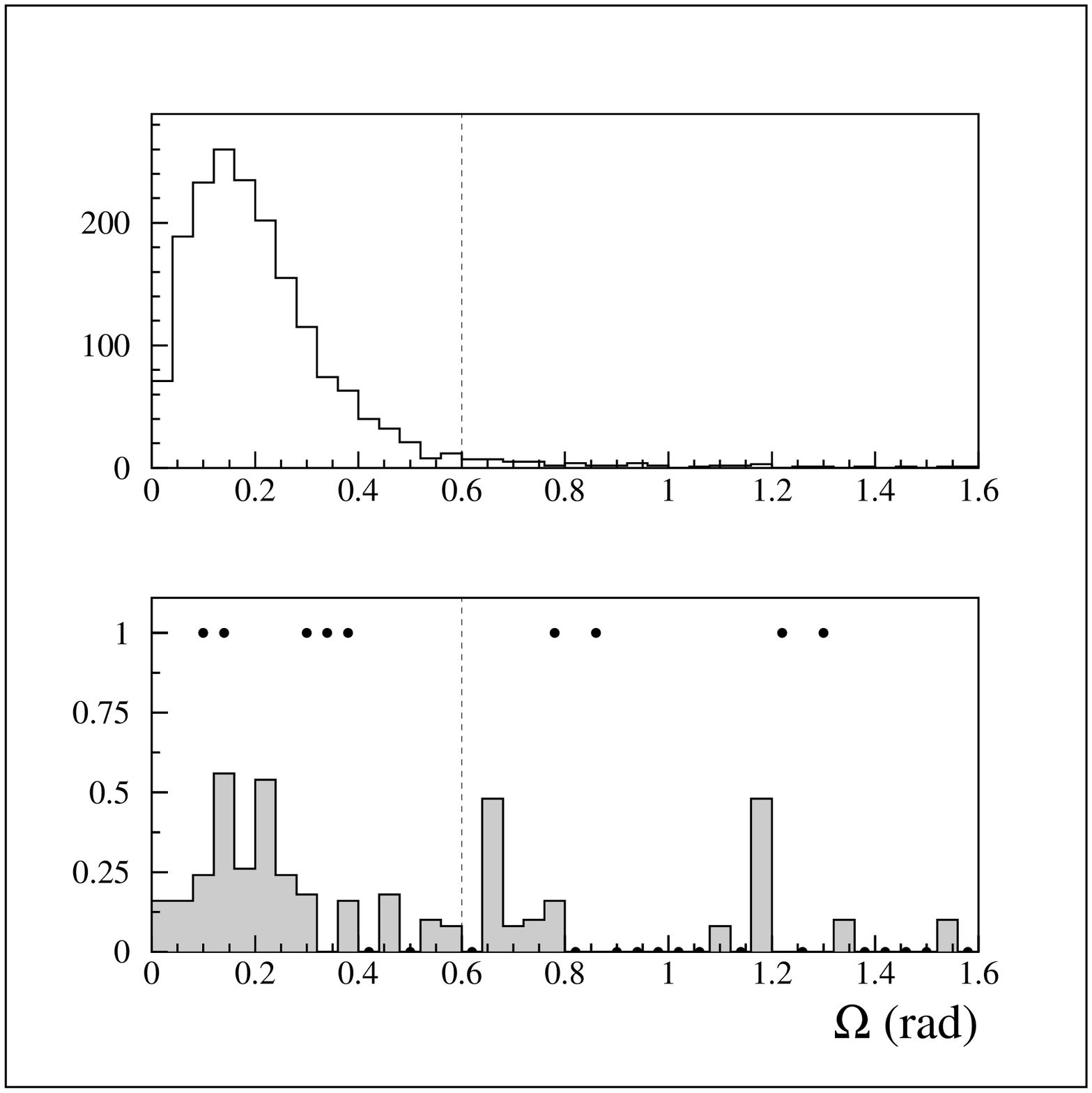,width=0.44\textwidth,clip=,bbllx=50pt,bblly=210pt,bburx=550pt,bbury=700pt}} 
  \end{center}
\vspace{1.cm}
  \caption{\small 
$\Omega$ angle between photon direction and missing momentum direction. 
Top: $\phi\rightarrow K^0\overline{K}^0\gamma$ MC. Bottom: data (dots) and background MC (grey).
Selection cuts are shown. Background MC samples are normalized to data taking into account the different LF values.}
   \label{fig:tetafot}
\end{figure}

We then look for the radiative photon, by searching for a calorimeter cluster, without any associated track.
The cluster must be in time with the observed event, and its position in the calorimeter must be compatible with the direction of the missing momentum. The angle $\Omega$ between the missing momentum and the photon direction, derived
from the cluster position, is required to be below 0.6 rad. Figure \ref{fig:tetafot} shows the $\Omega$ distributions for data and MC simulations.

 We verified the correct simulation of low-energy clusters in the MC using the $\phi \to \pi^+\pi^-\pi^0$ control sample: a time resolution of 520 ps is achieved for low energy clusters in both data and MC simulation, well reproduced by MC signal simulation.

\begin{figure}
  \begin{center}  
  \mbox{\epsfig{figure=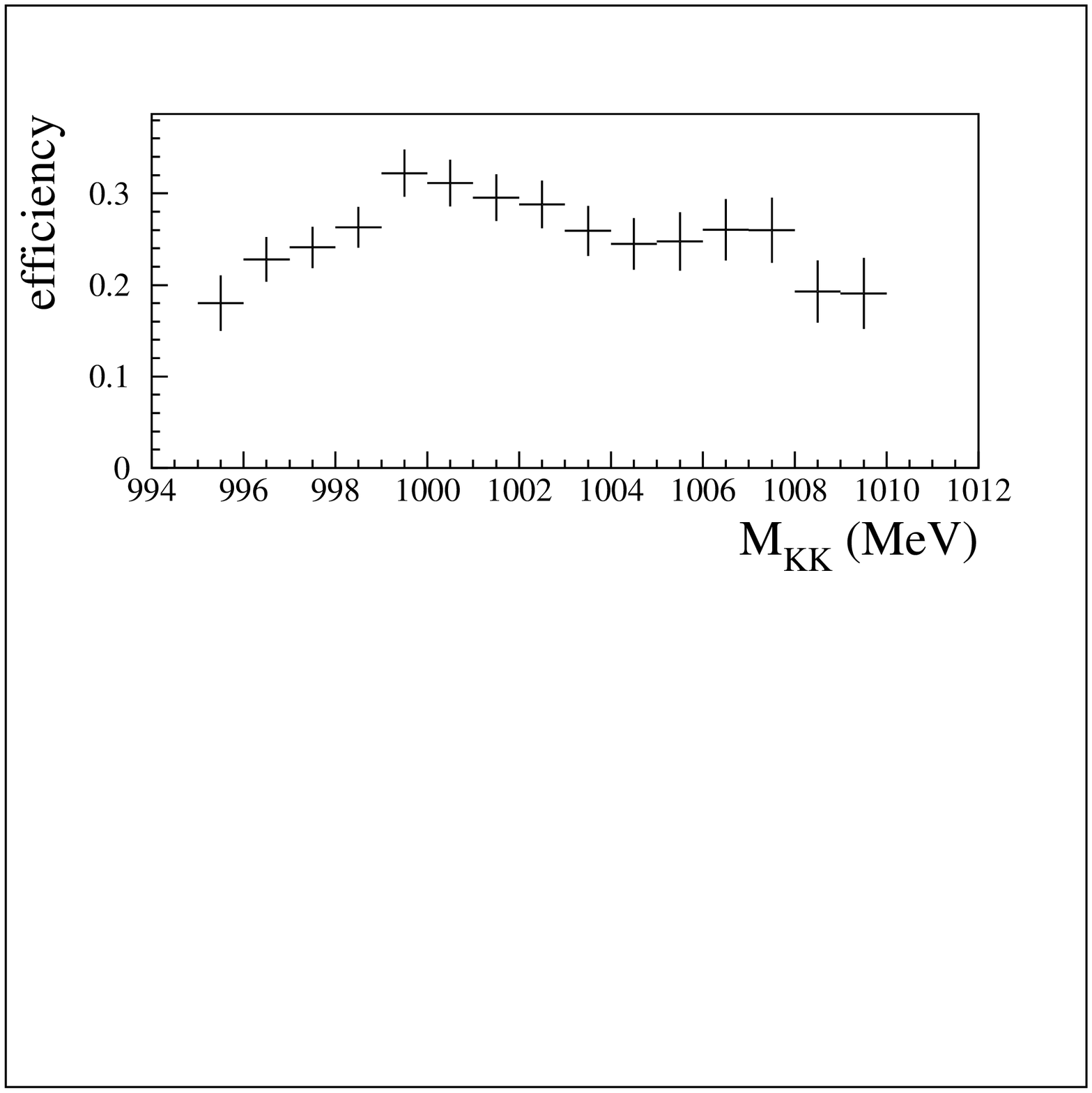,width=0.44\textwidth,clip=,bbllx=50pt,bblly=250pt,bburx=550pt,bbury=600pt}} 
  \end{center}
\vspace{-2.cm}
  \caption{\small Signal detection efficiency as a function of $M_{KK}$ in the allowed range.}
   \label{fig:effi_vs_mkk} 
\end{figure}

After the selection, 33 events are present in the non-rescaled MC samples for CP-violating rare decays and $e^+e^- \rightarrow \pi^+\pi^-\pi^+\pi^-(\gamma)$ decays. No event is present in the MC sample for all $\phi$ decays. The overall efficiency for the signal is (24.8$\pm$0.5)\%, evaluated from signal simulation. The efficiency is shown in fig.~\ref{fig:effi_vs_mkk} as a function of $M_{KK}$ .
The evaluation of this efficiency can vary depending on the scalar field dynamics and the size of couplings used for the signal simulation.
To study this effect the kaon-loop approach~\cite{Achasov2} has been used, 
varying the couplings $g_{SKK}$, with S=$f_{0},a_{0}$ in ranges larger than the present spread  
of experimental measurements and theoretical evaluations. Signal efficiency has been recalculated throughout this range and compared with the one used in the analysis: the variation is within the statistical error.

\begin{figure}
  \begin{center}  
   \mbox{ \epsfig{figure=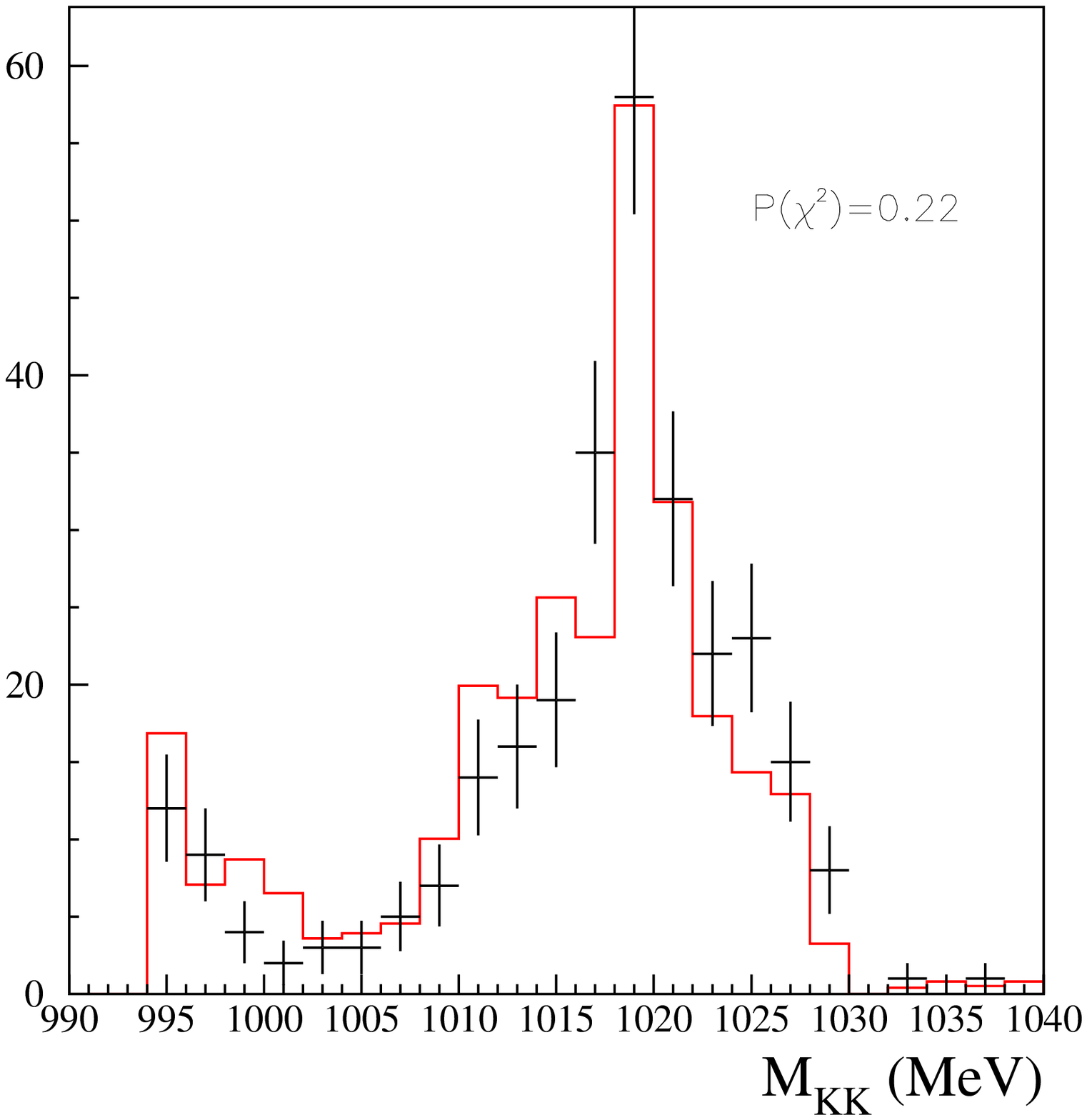,width=0.44\textwidth,clip=,bbllx=50pt,bblly=210pt,bburx=550pt,bbury=700pt}  } 
   \mbox{\epsfig{figure=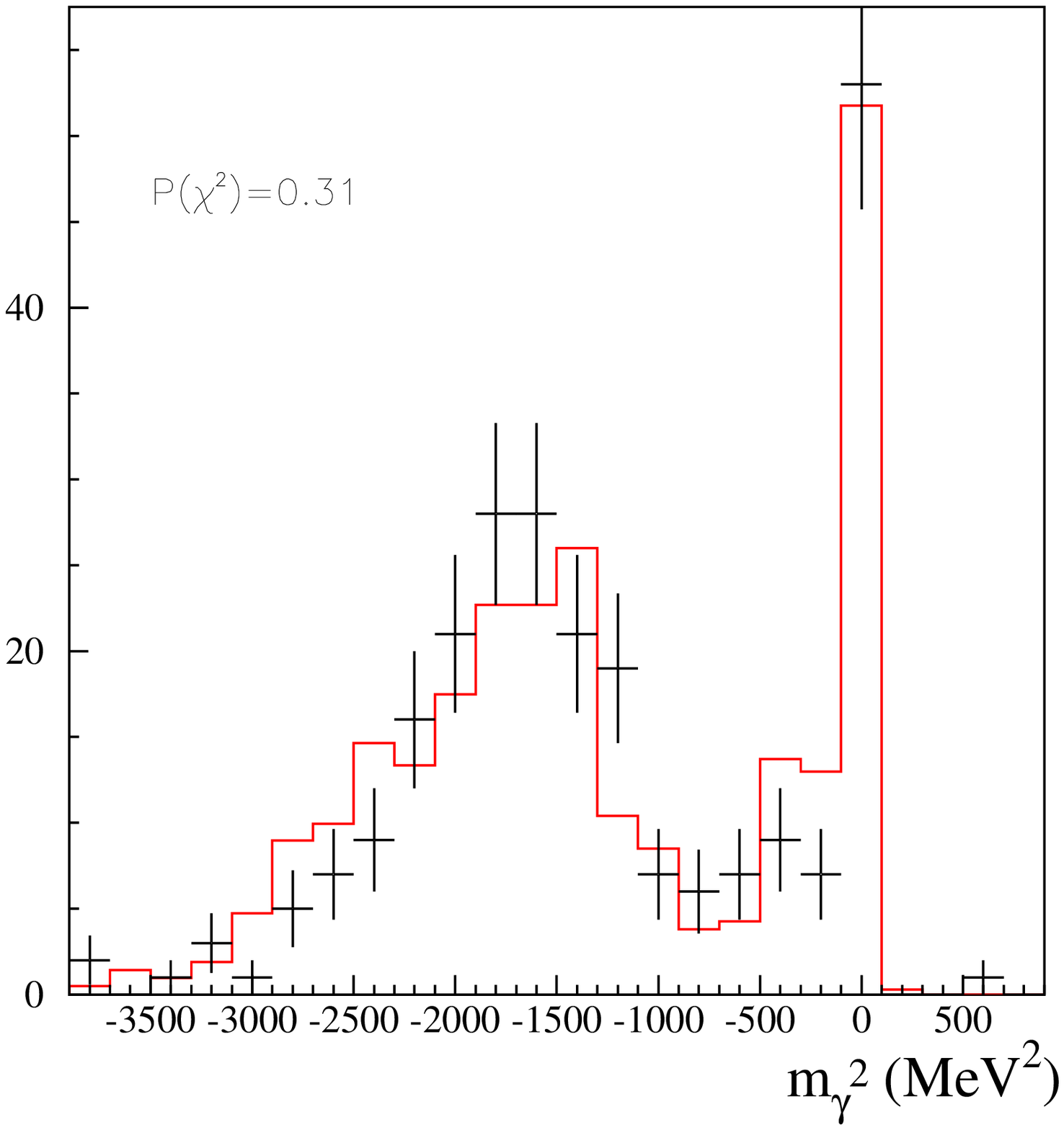,width=0.44\textwidth,clip=,bbllx=50pt,bblly=210pt,bburx=550pt,bbury=700pt}} 
  \end{center}
\vspace{1.cm}
  \caption{\small Data (points) and MC background fit result (solid) for $M_{KK}$ (left) and $m_{\gamma}^2$ (right) distributions.}
   \label{fig:fits}
\end{figure}

To normalize MC distributions to data we have estimated effective scale factors, one for each MC sample, fitting the MC distributions for M$_{KK}$ and $m_{\gamma}^2$ to the corresponding data distributions. The results of the fits are shown in fig.~\ref{fig:fits}. For this purpose we used a control sample made of events surviving all the selection requirements, except the M$_{KK}$ and $m_{\gamma}^2$ cuts. For the simulation of all $\phi$ decays, the scale factor obtained from the fit is equal to $0.47\pm0.04$, 
while the MC scale factor for CP-violating rare decays and $e^+e^- \rightarrow \pi^+\pi^-\pi^+\pi^-(\gamma)$ decays is $0.096\pm0.015$. These factors differ from the LF scale factors by at most 15\% and are in agreement within errors. Rescaled background events are listed in tab.~\ref{tab2}. 



The expected number of background events, once scale factors obtained from fits are applied, is 3.2$\pm$0.7 ; the error is the combination of the statistical error and the systematic error due to the fit. The number of residual events in the data sample is five. These numbers will be used for the upper limit evaluation.

\begin{table}[htbp]
\begin{center}
{\begin{tabular}{lcc}
\hline
\hline
 Background events & Rare kaon decays & $e^+e^- \rightarrow \pi^+\pi^-\pi^+\pi^-(\gamma)$ \\
\hline
non-rescaled& 26.0$\pm$5.1 & 7.0$\pm$2.6\\
rescaled by LF& 2.13$\pm$0.42 & 0.70$\pm$0.26\\
rescaled by fit result& 2.50$\pm$0.62 & 0.67$\pm$0.27 \\
\hline
\hline
\end{tabular} }
\vskip 0.4cm
\caption{Expected background events, before and after rescaling. Errors are the combination of both statistical and systematic contributions. Residual events obtained by rescaling by fit or LF are in agreement within the errors.
\label{tab2}} 
\end{center}
\end{table}

\section{Result}
The upper limit on BR($\phi\to K^0\overline{K}^0\gamma$) 
is evaluated as follows:
\begin{equation}
BR < \frac{S_{\rm{90CL}}}{\int{L dt}\times\sigma(e^+e^-\to \phi)\times1/2\times(BR(K_S\to\pi^+\pi^-))^2\times\epsilon} .
\end{equation}
$\int{L dt}$ is the integrated luminosity, 2.18 fb$^{-1}$.  
$\sigma(e^+e^-\to \phi)$ is the cross section for $\phi$ production from
$e^+e^-$ 
annihilation, equal to 3.09 $\mu$b
; this is evaluated from the fit to the visible cross section for the process $\phi \to \eta \gamma$ with the data taken around the $\phi$ peak, with a fractional error of 2.4\%. 
 $BR(K_S\to\pi^+\pi^-)$ is taken as 0.692~\cite{pdg2006}. 
The signal efficiency $\epsilon$ is $(24.8\pm0.5)\%$.
The upper limit on the number of events at 90\% CL, $S_{\rm{90CL}}$, is
calculated using the so-called 
``unified approach''
~\cite{feldcous}, using the number of residual data events, and the residual background events reduced by 1$\sigma$ to take into account the systematic error on the background evaluation in a conservative way;
 we get $S_{\rm{90CL}}$ = 7.49. 
Our result for the upper limit is:
\begin{equation}
BR(\phi \to K^0\overline{K}^0\gamma) < 1.9 \times 10^{-8} .
\end{equation}

\begin{figure}
  \begin{center}  
  \mbox{\epsfig{figure=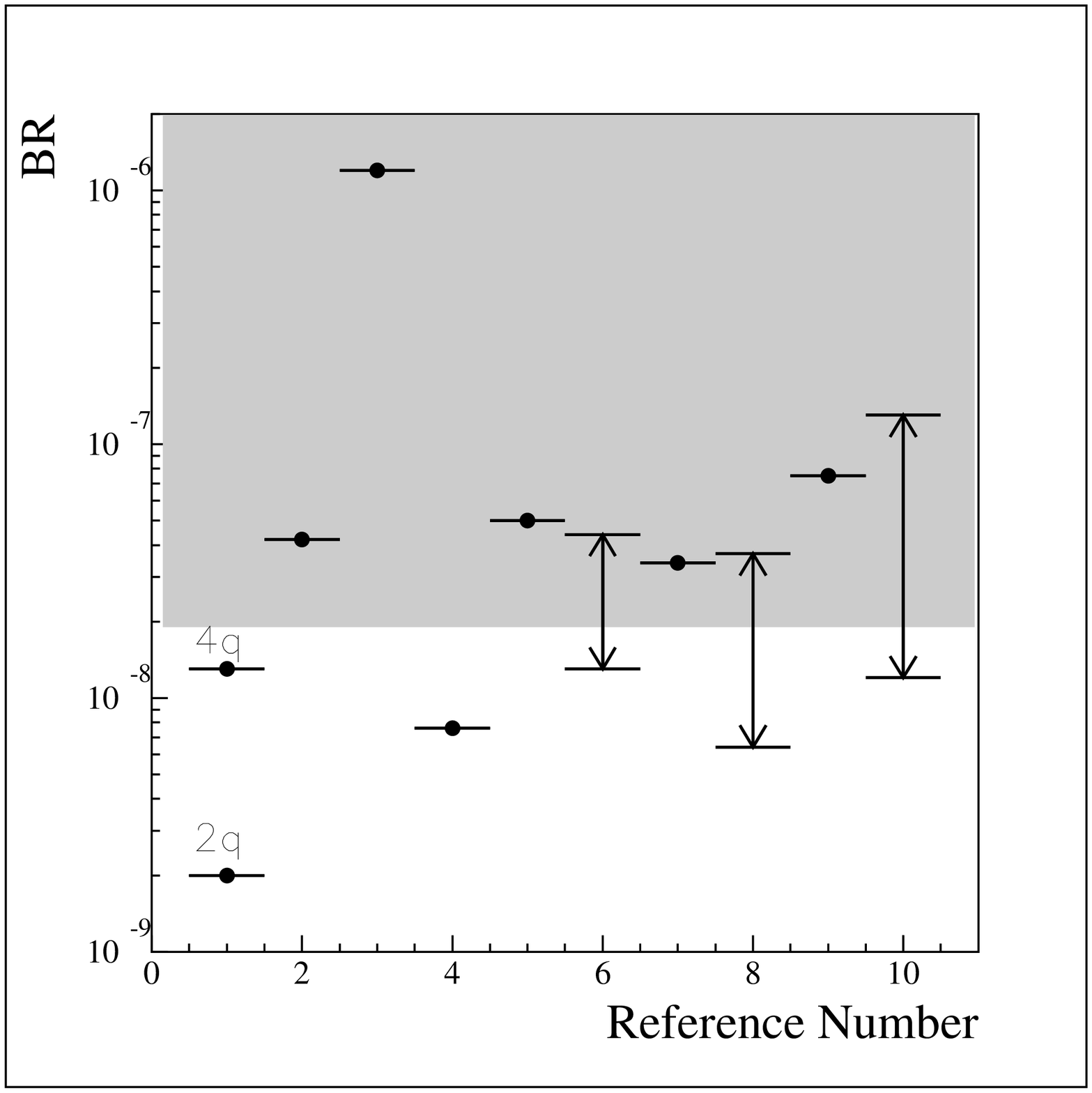,width=0.6\textwidth,clip=,bbllx=50pt,bblly=210pt,bburx=550pt,bbury=550pt}} 
  \end{center}
\vspace{1.cm}
  \caption{\small Comparison between the region excluded by the KLOE 90\% CL upper limit on $BR(\phi\to K^0\overline{K}^0\gamma)$ (hatched area) and theoretical estimates.           The numbers in abscissa are the reference numbers.}
   \label{fig:br_excluded} 
\end{figure}

This is the only existing experimental result for this branching ratio. In fig.~\ref{fig:br_excluded} the upper limit is compared with theoretical predictions~\cite{Achasov1,Fajfer,Lucio,Bramon2,Oller1,Achasov2,Nussinov,Oller2,Escribano,Gokalp}. 
Ref.~\cite{Achasov1} is based on the scalar kaon-loop model: in fig.~\ref{fig:br_excluded} $2q$ and $4q$ indicate the prediction for a scalar $q\bar{q}$ structure or a tetraquark structure respectively; 
Ref.~\cite{Fajfer} describes this decay with a low-energy effective lagrangian.
Ref.~\cite{Lucio} only takes into account the $f_0$ contribution, without considering the $a_0$, which explains the large BR value; the authors criticize the result of ref.~\cite{Nussinov}.
Ref.~\cite{Bramon2} makes use of a chiral perturbation lagrangian which does not include explicitly the scalar mesons in the intermediate state.
Ref.~\cite{Oller1,Oller2} calculates this BR through unitarized chiral perturbation theory based on resonant and non-resonant $\phi \to (f_0+a_0)$ coupling. 
In ref.~\cite{Achasov2} the couplings of the kaon-loop model are evaluated by using the SND experiment data~\cite{achadata}.
Ref.~\cite{Escribano} introduces a well-defined $U(3) \times U(3)$ chiral model which incorporates {\it ab initio} the
pseudoscalar and scalar mesons nonets.
Ref.~\cite{Gokalp} repeats the calculation of ref.~\cite{Achasov2} using the first published KLOE data~\cite{pl1,pl2}, with and without the $\sigma$(600) contribution, and gives a large set of values.

Following the kaon-loop model~\cite{Achasov2} we have evaluated predictions for $BR(\phi\to K^0\overline{K}^0\gamma)$ based on the $g_{f0KK}$, $g_{f0\pi\pi}$ and $g_{a0KK}$ couplings obtained by KLOE~\cite{pl1,pl2,pl3} by fitting the data of $\phi$ decays into $\pi^+ \pi^- \gamma$, $\pi^0 \pi^0 \gamma$ and $\eta \pi \gamma$.
Allowing for variations of the couplings within the corresponding errors, the predicted BRs vary between 0.4$\times 10^{-8}$ and 6.8$\times 10^{-8}$, the lower values being still compatible with the present upper limit.


\section{Acknowledgements}

We thank the DA$\Phi$NE team for their efforts in maintaining low background running 
conditions and their collaboration during all data taking. 
We want to thank our technical staff: 
G.F. Fortugno and F. Sborzacchi for their dedicated work to ensure an
efficient operation of 
the KLOE computing facilities; 
M. Anelli for his continuous support to the gas system and the safety of
the
detector; 
A.  Balla, M. Gatta, G. Corradi and G. Papalino for the maintenance of the
electronics;
M. Santoni, G. Paoluzzi and R. Rosellini for the general support to the
detector; 
C. Piscitelli for his help during major maintenance periods.
This work was supported in part
by EURODAPHNE, contract FMRX-CT98-0169; 
by the German Federal Ministry of Education and Research (BMBF) contract 06-KA-957; 
by the German Research Foundation (DFG),'Emmy Noether Programme',
contracts DE839/1-4;
and by the EU Integrated
Infrastructure
Initiative HadronPhysics Project under contract number
RII3-CT-2004-506078.


\begin{thebibliography}{99}
\bibitem{Achasov1} N.N.~Achasov, V.N.~Ivanchenko, Nucl. Phys. B315 (1989) 465. 
\bibitem{Fajfer} S.~Fajfer, R.J.~Oakes, Phys. Rev. D42 (1990) 2392.
\bibitem{Lucio} J.~Lucio, J.~Pestieau, Phys. Rev. D42 (1990) 3253.
\bibitem{Bramon2} A.~Bramon, A.~Grau, G.~Pancheri, Phys. Lett. B289 (1992) 97.
\bibitem{Oller1} J.A.~Oller, Phys. Lett. B426 (1998) 7.
\bibitem{Achasov2} N.N.~Achasov, V.V.~Gubin, Phys. Rev. D64 (2001) 094016. 
\bibitem{Nussinov} S.~Nussinov, T.N.~Truong, Phys. Rev. Lett. 63 (1989) 1349, Erratum-ibid, {\bf 63} (1989) 2002. 
\bibitem{Oller2} J.A.~Oller, Nucl. Phys. A714 (2003) 161.
\bibitem{Escribano} R.~Escribano, Eur. Phys. J. A31 (2007) 454.
\bibitem{Gokalp} A.~Gokalp, C.S.~Korkmaz, O.~Yilmaz, Phys. Rev. D75 (2007) 013001.
\bibitem{feldcous} G.J.~Feldman, R.D.Cousins, Phys. Rev.  D57 (1998) 3873.
\bibitem{dc} M.~Adinolfi et al. (KLOE Collaboration), Nucl. Instr. Meth. A 488 (2002) 51.
\bibitem{emc} M.~Adinolfi et al. (KLOE Collaboration), Nucl. Instr. Meth. A 482 (2002) 364.
\bibitem{trigger}M.~Adinolfi et al. (KLOE Collaboration), Nucl. Instr. Meth. A 492 (2002) 134.
\bibitem{lumbha} F.~Ambrosino et al. (KLOE Collaboration), Eur. Phys. J. C47 (2006) 589-596.
\bibitem{OFFLINE} F.~Ambrosino et al. (KLOE Collaboration), Nucl. Instr. Meth. A 534 (2004) 403.
\bibitem{4pcmd2} B.~Aubert et al. (BaBar Collaboration), Phys. Rev. D71 (2005) 052001.

\bibitem{rodrigo} G. Rodrigo, Nucl. Phys.  Proc. Suppl. 169 (2007) 271.
\bibitem{pdg2006} C. Amsler et al., Phys. Lett. B667, 1 (2008).
\bibitem{helene} O.~Helene, Nucl. Instr.  and Meth.  212 (1983) 319.
\bibitem{pl1} A.~Aloisio et al. (KLOE Collaboration), Phys. Lett. B536 (2002) 209.
\bibitem{pl2} A.~Aloisio et al. (KLOE Collaboration), Phys. Lett. B537 (2002) 21.
\bibitem{pl3} F.~Ambrosino et al. (KLOE Collaboration), Phys. Lett. B634 (2006) 14.
\bibitem{achadata} N.N.~Achasov, V.V.~Gubin, Phys. Rev. D63 (2001) 094007.
\end{thebibliography}
\end{document}